%%%%%%%If you do not have the msbm fonts, delete the following 4 lines
\font\mybb=msbm10 at 12pt
\def\bbbb#1{\hbox{\mybb#1}}
\def\Z {\bbbb{Z}}
\def\R {\bbbb{R}}
%%%%%%%%%%%%
%%%and replace with the following 2 lines (without %)
%\def\Z {Z}
%\def\R {R}
%%%%%%%%%%
%\def \r11 {{R_{11}}
\def \tti {{$\ti T^d \times \R$}}

\def \aa {\alpha}
\def \bb {\beta} 
\def \gg {\gamma}

\def \pp {\pi}

\def \ss {\sigma}

\def \ww{\omega}

\def \ti {\tilde}

\def \2 {{1 \over 2}}
\def \3 {{1 \over 3}}
\def \4 {{1 \over 4}}
\def \5 {{1 \over 5}}
\def \6 {{1 \over 6}}
\def \7 {{1 \over 7}}
\def \8 {{1 \over 8}}
\def \9 {{1 \over 9}}
\def \0 { \infty}

\def\ek {\eqn\abc$$}

  %%%%%%%This requires the PHYZZX.TEX macropackage

\tolerance=10000
\input phyzzx
%%%%%%%%%%%%%%%%%%%%%%%%%%%%%%%%%%%%%%%%%%%%%%%%%%%%%%%%%%%%%%%%%%%%

 \def\unit{\hbox to 3.3pt{\hskip1.3pt \vrule height 7pt width .4pt \hskip.7pt
\vrule height 7.85pt width .4pt \kern-2.4pt
\hrulefill \kern-3pt
\raise 4pt\hbox{\char'40}}}

\def \pl {Phys. Lett. B}
\def  \np {Nucl. Phys. B}
%%%%%%%%%%%%%%%%%%%%%%%%%%%%%%%%%%%%%%%%%%%%%%%%%%%%%%%%%%%%%%%%%%%%

%%%%%%%%%%%%%%%%%%%%%%%%%%%%%%%%%%%%%%%%%%%%%%%%%%%%%%%%%%%%%%%%%%%%
%%%%%%%%%%%%%%%%%%%%%%%%%%%%%%%%%%%%%%%%%%%%%%%%%%%%%%%%%%%%%%%%%%%%
\REF\Mat{T. Banks, W. Fischler, S. Shenker, and L. Susskind,
``M theory as a matrix model: a conjecture,''
hep-th/9610043, Phys. Rev. 
{\bf D55} (1997) 5112.}%
\REF\brev{For a review see T. Banks,
hep-th/9706168, hep-th/9710231.}%
\REF\Sus{L. Susskind, hep-th/9704080.}
\REF\taylor{W. Taylor, hep-th/9611042, Phys. Lett. {\bf B394} (1997) 283.}%
\REF\motl{L. Motl,hep-th/9701025.}
\REF\IIAbs{T. Banks and N. Seiberg,hep-th/9702187.}
\REF\IIAVV{R. Dijkgraaf,E. Verlinde and H. Verlinde, hep-th/9703030.}
\REF\dvv{R. Dijkgraaf, E. Verlinde, and H. Verlinde,  hep-th/9603126;
 hep-th/9604055; hep-th/9703030; hep-th/9704018.} 
\REF\rozali{M. Rozali, hep-th/9702136.}
\REF\brs{M. Berkooz, M. Rozali, and N. Seiberg, 
%``Matrix 
%description of M-theory on $T^4$ and $T^5$, 
hep-th/9704089}
\REF\fhrs{W. Fischler, E. Halyo, A. Rajaraman and L. Susskind, 
hep-th/9703102.}%
\REF\MatTor{N. Seiberg, 
%``New Theories in Six Dimensions and
%Matrix Description of M-theory on $T^5$ and $T^5/\Z_2$'',
hep-th/9705221.}%
\REF\Hv{F. Hacquebord and H. Verlinde, hep-th/9707179.}
\REF\Alg{S. Elitzur, A. Giveon, D. Kutasov and E. Rabinbovici, hep-th/9707217.}
\REF\mem{A. Losev, G. Moore and S.L. Shatashvili, hep-th/9707250; I. Brunner and A. Karch,
hep-th/9707259; A. Hanany and G. Lifschytz, hep-th/9708037; R. Argurio and L. Houart, hep-th/9710027.}
\REF\Sen{A. Sen, hep-th/9709220.}
\REF\Seib{N. Seiberg, 
hep-th/9710009.}%
\REF\Bol{M. Blau and M. O'Loughlin, hep-th/9712047.}
\REF\progj{C.M. Hull and B. Julia, in preparation}
\REF\HT{C.M. Hull and P.K. Townsend, hep-th/9410167.}
\REF\town{P.K. Townsend, hep-th/9501068.}
\REF\GravDu{C.M. Hull, hep-th/9705162.}
\REF\StrMem{C.M. Hull, hep-th/9512181.}
\REF\humat{C.M. Hull, hep-th/9711179.}
\REF\doug{A. Connes, M.R. Douglas and A. Schwarz, hep-th/9711162;
M.R. Douglas and C.M. Hull, hep-th/9711165.}

%%%%%%%%%%%%%%%%%%%%%%%%%%%%%%%%%%%%%%%%%%%%%%%%%%%%%%%%%%%%%%%%%%%%
%%%%%%%%%%%%%%%%%%%%%%%%%%
%%@@

%%%%%%%%%%%%%%%%%%%%%%%%%%%%%%%%
%
% S-Tables Macro
%
%\message{S-Tables Macro v1.0, ACS, TAMU (RANHELP@VENUS.TAMU.EDU)}
%
% Help Text
%
\newhelp\stablestylehelp{You must choose a style between 0 and 3.}%
\newhelp\stablelinehelp{You should not use special hrules when stretching
a table.}%
\newhelp\stablesmultiplehelp{You have tried to place an S-Table inside another
S-Table.  I would recommend not going on.}%
%
% Line Thicknesses (Values)
%
\newdimen\stablesthinline
\stablesthinline=0.4pt
\newdimen\stablesthickline
\stablesthickline=1pt
%
% Border and Internal Line Thicknesses
%
\newif\ifstablesborderthin
\stablesborderthinfalse
\newif\ifstablesinternalthin
\stablesinternalthintrue
\newif\ifstablesomit
\newif\ifstablemode
\newif\ifstablesright
\stablesrightfalse
%
% Save Registers
%
\newdimen\stablesbaselineskip
\newdimen\stableslineskip
\newdimen\stableslineskiplimit
%
% Counters
%
\newcount\stablesmode
\newcount\stableslines
\newcount\stablestemp
\stablestemp=3
\newcount\stablescount
\stablescount=0
\newcount\stableslinet
\stableslinet=0
%
% Table Style Selection
%
% 0 - Centered
% 1 - Left Justified
% 2 - Right Justified
% 3 - Not Justified
%
\newcount\stablestyle
\stablestyle=0
%
% Element Buffering Definitions
%
\def\stablesleft{\quad\hfil}%
\def\stablesright{\hfil\quad}%
%
% Vertical Bar Activation
%
\catcode`\|=\active%
%
% Strut Control
%
\newcount\stablestrutsize
\newbox\stablestrutbox
\setbox\stablestrutbox=\hbox{\vrule height10pt depth5pt width0pt}
\def\stablestrut{\relax\ifmmode%
                         \copy\stablestrutbox%
                       \else%
                         \unhcopy\stablestrutbox%
                       \fi}%
%
% Misc. Internal Stuff
%
\newdimen\stablesborderwidth
\newdimen\stablesinternalwidth
\newdimen\stablesdummy
\newcount\stablesdummyc
\newif\ifstablesin
\stablesinfalse
%
% Table Macros
%
\def\begintable{\stablestart%
  \stablemodetrue%
  \stablesadj%
  \halign%
  \stablesdef}%
\def\stablesadj{%
  \ifcase\stablestyle%
    \hbox to \hsize\bgroup\hss\vbox\bgroup%
  \or%
    \hbox to \hsize\bgroup\vbox\bgroup%
  \or%
    \hbox to \hsize\bgroup\hss\vbox\bgroup%
  \or%
    \hbox\bgroup\vbox\bgroup%
  \else%
    \errhelp=\stablestylehelp%
    \errmessage{Invalid style selected, using default}%
    \hbox to \hsize\bgroup\hss\vbox\bgroup%
  \fi}%
\def\stablesend{\egroup%
  \ifcase\stablestyle%
    \hss\egroup%
  \or%
    \hss\egroup%
  \or%
    \egroup%
  \or%
    \egroup%
  \else%
    \hss\egroup%
  \fi}%
\def\stablestart{%
  \ifstablesin%
    \errhelp=\stablesmultiplehelp%
    \errmessage{An S-Table cannot be placed within an S-Table!}%
  \fi
  \global\stablesintrue%
  \global\advance\stablescount by 1%
  \message{<S-Tables Generating Table \number\stablescount}%
  \begingroup%
  \stablestrutsize=\ht\stablestrutbox%
  \advance\stablestrutsize by \dp\stablestrutbox%
  \ifstablesborderthin%
    \stablesborderwidth=\stablesthinline%
  \else%
    \stablesborderwidth=\stablesthickline%
  \fi%
  \ifstablesinternalthin%
    \stablesinternalwidth=\stablesthinline%
  \else%
    \stablesinternalwidth=\stablesthickline%
  \fi%
  \tabskip=0pt%
  \stablesbaselineskip=\baselineskip%
  \stableslineskip=\lineskip%
  \stableslineskiplimit=\lineskiplimit%
  \offinterlineskip%
  \def\borderrule{\vrule width \stablesborderwidth}%
  \def\internalrule{\vrule width \stablesinternalwidth}%
  \def\thinline{\noalign{\hrule height \stablesthinline}}%
  \def\thickline{\noalign{\hrule height \stablesthickline}}%
  \def\trule{\omit\leaders\hrule height \stablesthinline\hfill}%
  \def\ttrule{\omit\leaders\hrule height \stablesthickline\hfill}%
  \def\tttrule##1{\omit\leaders\hrule height ##1\hfill}%
  \def\stablesel{&\omit\global\stablesmode=0%
    \global\advance\stableslines by 1\borderrule\hfil\cr}%
  \def\el{\stablesel&}%
  \def\elt{\stablesel\thinline&}%
  \def\eltt{\stablesel\thickline&}%
  \def\elttt##1{\stablesel\noalign{\hrule height ##1}&}%
  \def\elspec{&\omit\hfil\borderrule\cr\omit\borderrule&%
              \ifstablemode%
              \else%
                \errhelp=\stablelinehelp%
                \errmessage{Special ruling will not display properly}%
              \fi}%
  \def\stmultispan##1{\mscount=##1 \loop\ifnum\mscount>3 \stspan\repeat}%
  \def\stspan{\span\omit \advance\mscount by -1}%
  \def\multicolumn##1{\omit\multiply\stablestemp by ##1%
     \stmultispan{\stablestemp}%
     \advance\stablesmode by ##1%
     \advance\stablesmode by -1%
     \stablestemp=3}%
  \def\multirow##1{\stablesdummyc=##1\parindent=0pt\setbox0\hbox\bgroup%
    \aftergroup\emultirow\let\temp=}
  \def\emultirow{\setbox1\vbox to\stablesdummyc\stablestrutsize%
    {\hsize\wd0\vfil\box0\vfil}%
    \ht1=\ht\stablestrutbox%
    \dp1=\dp\stablestrutbox%
    \box1}%
  \def\stpar##1{\vtop\bgroup\hsize ##1%
     \baselineskip=\stablesbaselineskip%
     \lineskip=\stableslineskip%
     \lineskiplimit=\stableslineskiplimit\bgroup\aftergroup\estpar\let\temp=}%
  \def\estpar{\vskip 6pt\egroup}%
  \def\stparrow##1##2{\stablesdummy=##2%
     \setbox0=\vtop to ##1\stablestrutsize\bgroup%
     \hsize\stablesdummy%
     \baselineskip=\stablesbaselineskip%
     \lineskip=\stableslineskip%
     \lineskiplimit=\stableslineskiplimit%
     \bgroup\vfil\aftergroup\estparrow%
     \let\temp=}%
  \def\estparrow{\vfil\egroup%
     \ht0=\ht\stablestrutbox%
     \dp0=\dp\stablestrutbox%
     \wd0=\stablesdummy%
     \box0}%
  \def|{\global\advance\stablesmode by 1&&&}%
  \def\|{\global\advance\stablesmode by 1&\omit\vrule width 0pt%
         \hfil&&}%
  \def\vt{\global\advance\stablesmode by 1&\omit\vrule width \stablesthinline%
          \hfil&&}%
  \def\vtt{\global\advance\stablesmode by 1&\omit\vrule width
\stablesthickline%
          \hfil&&}%
  \def\vttt##1{\global\advance\stablesmode by 1&\omit\vrule width ##1%
          \hfil&&}%
  \def\vtr{\global\advance\stablesmode by 1&\omit\hfil\vrule width%
           \stablesthinline&&}%
  \def\vttr{\global\advance\stablesmode by 1&\omit\hfil\vrule width%
            \stablesthickline&&}%
  \def\vtttr##1{\global\advance\stablesmode by 1&\omit\hfil\vrule width ##1&&}%
  \stableslines=0%
  \stablesomitfalse}
\def\stablesdef{\bgroup\stablestrut\borderrule##\tabskip=0pt plus 1fil%
  &\stablesleft##\stablesright%
  &##\ifstablesright\hfill\fi\internalrule\ifstablesright\else\hfill\fi%
  \tabskip 0pt&&##\hfil\tabskip=0pt plus 1fil%
  &\stablesleft##\stablesright%
  &##\ifstablesright\hfill\fi\internalrule\ifstablesright\else\hfill\fi%
  \tabskip=0pt\cr%
  \ifstablesborderthin%
    \thinline%
  \else%
    \thickline%
  \fi&%
}%
\def\endtable{\advance\stableslines by 1\advance\stablesmode by 1%
   \message{- Rows: \number\stableslines, Columns:  \number\stablesmode>}%
   \stablesel%
   \ifstablesborderthin%
     \thinline%
   \else%
     \thickline%
   \fi%
   \egroup\stablesend%
\endgroup%
\global\stablesinfalse}
%
% end of STABLES.TEX
%

%\input /LocalApps/TeXTables.app/stables.tex

%%%%%%%%%

%\input /LocalApps/TeXTables.app/stables.tex

%\catcode`\|=12

%\end

\vskip 1cm

%%%%
%\input /LocalApps/TeXTables.app/stables.tex

%\catcode`\|=12

%\end

%\catcode`\|=12

%%%%%%%%%%%%%%%%%%%%%%%%%%%%%%%%%%%%%%%%%%%%%%%%%%%%%%%%%%%%%%%%%%%%
\Pubnum{ \vbox{  \hbox {QMW-97-39} \hbox{LPTENS 97/59}\hbox{hep-th/9712075}} }
\pubtype{}
\date{December, 1997}

\titlepage

\title {\bf   U-Duality and BPS Spectrum of Super Yang-Mills Theory and M-Theory}

\author{C.M. Hull}
\address{Physics Department, Queen Mary and Westfield College,
\break Mile End Road, London E1 4NS, U.K.}
\andaddress{Laboratoire de Physique Th\' eorique, 
\break Ecole Normale Sup\' erieure, 
24 Rue Lhomond, 75231 Paris Cedex 05, France.}

\vskip 0.5cm

\abstract {
It is shown that the BPS spectrum of Super-Yang-Mills theory on $T^d\times \R$,
which  
 fits into
representations of the U-duality group for M-theory compactified on $T^{d}$, in accordance with the
matrix-theory conjecture, in fact 
 fits into
representations of the U-duality group for M-theory   on $T^{d+1}$, with the extra dualities
realised as generalised Nahm   transformations. 
The spectrum of BPS M-branes   is analysed,   new branes are discussed and matrix theory
applications described. }

\endpage

\chapter {Introduction}

A remarkable relation has been proposed between Super-Yang-Mills (SYM) theory and M-theory
[\Mat-\Seib]. M-theory in the infinite momentum frame on $R^{10}\times S^1$, in the limit in which
the radius of the \lq longitudinal' circle becomes infinite, $R_{11} \to \infty$, corresponds to one
dimensional  $U(N)$ SYM in the limit $N \to \infty$.   M theory compactified on a $d$ torus, i.e. on
$T^d\times R^{10-d}\times S^1$, in the limit  
 $R_{11} \to \infty$, corresponds to a theory whose low-energy limit is
the  $N \to \infty$ limit of
$d+1$ dimensional  $U(N)$ SYM 
on $\ti T^d \times \R$ where $\ti T^d$ is the dual torus and $\R$ is the time direction
[\Mat-\Seib]. The $d+1$ dimensional SYM is given by dimensional reduction of 10-dimensional SYM.
For $d\le 3$, the SYM theory is a consistent quantum theory and gives the full matrix theory description of the
M-theory. For $d>3$, however, extra degrees of freedom become important at short distances and
strong coupling, so that SYM is only an effective theory. For $d=4$, the full theory is the $5+1$
dimensional  (2,0) tensor multiplet theory [\rozali-\MatTor],  and 
for $d=5$ it is a $5+1$ dimensional non-critical string theory
 [\dvv-\MatTor].
In particular,   M-theory compactified on a $d$ torus has U-duality group 
$E_d(\Z)$ [\HT] (where  $E_5=SO(5,5)$, $E_4=SL(5)$,
$E_3=SL(3)\times SL(2)$,  $E_2=SL(2)\times \R$ and $E_d=E_{d,d}$ for $d=6,7,8$), and this is a manifest
symmetry of the matrix
 theory, at least for $d\le 5$. The $SL(d,\Z)$ subgroup of  $E_d(\Z)$
is the geometrical symmetry associated with the torus $\ti T^d$. For $d=3$, the 
duality group is
$SL(3)\times SL(2)$, and the extra $SL(2,\Z)$ arises from the S-duality of 3+1 dimensional SYM.

It has  been proposed that M-theory on $T^d\times R^{10-d}\times S^1$ 
for {\it finite} $R_{11}$, in the discrete light cone gauge,
is related to $U(N)$ SYM on $\ti T^d \times \R$ for {\it finite} $N$ [\Sus].
The case in which the extra   circle is null can be
thought of as an infinite boost  limit of the space-like reduction [\Sen,\Seib], and the fact that
  M-theory compactified on $T^d\times S^1$ 
where the extra circle is space-like 
has a U-duality group $E_{d+1} $  suggests that 
 the $d+1$ dimensional 
SYM on $\ti T^d\times \R$ should have $E_{d+1}(\Z)$ (or perhaps a discrete subgroup of a
contraction of $E_{d+1}$), and should   have more than just the
   the  $E_{d }(\Z)$ symmetry of
the large $N$ limit. In [\progj], 
the duality groups for null reductions of 11-dimensional supergravity will be considered.
  In [\Hv], evidence was given that for $d=3$ there is $E_4$ symmetry instead of the expected
$E_3$. For finite
$N$, it was argued that the expected
$E_3=SL(3)\times SL(2)$ symmetry is enhanced to $SL(5)$, with the extra duality symmetries
realised as Nahm-type transformations  that interchange the rank $N$ of the gauge group with electric
or magnetic fluxes. It was shown that the $E_3$ representations of BPS states fit into
representations of $E_4$.

In this paper, evidence will be given that
 the BPS spectrum of SYM indeed fits into representations of $E_{d+1}$ for other
values of $d$ (if a matrix theory exists for such $d$), based on and generalising the results of [\Hv,\Alg]. 
Similar results have also been
obtained by Blau and O'Loughlin [\Bol]. 
The information about BPS states so obtained will then  be used to re-examine the brane spectrum of
M-theory.
Indeed, much of the analysis here also applies to M-theory compactification on a conventional
(space-like) torus. Starting from the known 0-brane spectrum (from wrapped M2 and M5 branes, and
from pp-waves etc) and acting with U-duality,  the expected multiplets are obtained for
compactification to 5 dimensions or higher, but  for lower dimensions new exotic branes are
generated. In four dimensions, the extra branes are just the wrapped Kaluza-Klein (KK) monopole
6-branes, but for 3 dimensions, many extra branes are needed to complete the 0-brane multiplet,
which is found to be a {\bf 248} of $E_8$. Some of the extra branes were anticipated in [\GravDu]
and references therein, while others were first proposed in [\Alg].

For $d \ge 6$, it is not known whether a matrix theory limit exists, or whether there is a
consistent quantum theory with a $d+1$ dimensional SYM effective low-energy description, and the
results here will only apply for those dimensions in which these theories do exist.
For $d=3$, SYM on $T^3\times \R$ has $SL(3;\Z)$ torus symmetry, together with a conjectured
S-duality and conjectured duality under Nahm transformations, generating an $SL(5;\Z)$ symmetry
[\Hv]. The theory on $T^d\times \R$, if it exists, should reduce to 
SYM on $T^3\times \R$ at low energies and in any limit in which a $T^{d-3}$ shrinks, leaving an
effective theory with 
approximate S-duality and Nahm-duality symmetries.
In [\Alg], it was shown that if the S-duality of this limit extends to an exact symmetry of the
full $d+1$ dimensional theory, then the S-dualities together with the $SL(d;\Z)$ torus symmetries
together generate the U-duality group $E_d(\Z)$. Here, this will be extended to  show that if the
Nahm dualities also extend to $d+1$ dimensional exact symmetries, then $E_{d+1}(\Z)$ is generated.

In [\Sus,\Sen,\Seib]], the $d+1$ $U(N)$ SYM theory   arises from
the discrete light cone gauge quantization of M-theory on the product of $T^d$ with a null circle,
for fixed quantized longitudinal momentum, which is proportional to $N$. Including all possible
momenta, which is required for Lorentz invariance, requires the consideration of a larger theory,
with a quantum number $N$, such that the sector with a given $N$ has a $U(N)$ SYM description.
Moreover, an $N=0$ sector should presumably be included, even though it would not have a
conventional SYM description.

The Nahm dualities act in SYM by interchanging the rank $N$ with electric and magnetic fluxes, so
that it relates BPS states of a $U(N)$ theory with flux $N'$ to states of a dual $U(N')$ theory with
flux
$N$. It is not a symmetry of a given SYM theory, but could be a symmetry of the larger \lq matrix'
theory which includes sectors for all values of $N$. It will also include transformations relating
$N\ne 0$ sectors with the $N=0$ sector. 

The Nahm dualities   act in   M-theory by interchanging the longitudinal radius $R_{11}$ with  the
radius $R_i$  of a space-like circle, and so are relevant to the Lorentz invariance of the theory.
The U-dualities also act on the supergravity background fields, which correspond to torus moduli or
to generalised coupling constants of the SYM theory [\humat]. In particular, the Nahm
transformations in
$E_{d+1}(\Z)/E_d(\Z)$ generate non-trivial anti-symmetric tensor gauge-fields on the
null torus and for such  backgrounds, non-commutative geometry plays a role [\doug].
These situations are best described by SYM on a non-commutative torus [\doug], and a deeper
understanding of such situations is clearly needed. In the following, we will ignore such
considerations and preceed formally. 

\chapter{U-duality and the BPS Spectrum of SYM}

 We now review some of the results of [\Alg], in which the BPS spectrum of SYM
on
$\ti T^d
\times \R$ was shown to fit into representations of
$E_d$ (with  some subtleties   for $d \ge 7$).
We follow the notation of [\Alg]; $\ti T^d$ is taken to be  a 
rectangular torus with circles of radii $s_i$, $i=1,\dots , d$,
and the SYM coupling is $g$. This corresponds to M-theory on a
 rectangular torus $T^d\times  S^1$, with circles
of radii $R_i, R_{11}$, and coupling given by the $D=11$ Planck length $l_p$.
The volumes of the $d$-tori $\ti T^d,  T^d$ are
$$V_R= \prod_{i=1}^d R_i, \qquad V_s= \prod_{i=1}^d s_i
\eqn\abc$$
The relationship between the two sets of variables is
$$\eqalign{
s_i&={l_p^3
\over R_{11} R_i}
\cr
g^2&={l_p^{3(d-2)}
\over R_{11}V_R}
\cr}
\eqn\abc$$

The SYM theory on \tti\ is invariant under the manifest $SL(d,\Z)$, and this
is reduced to the permutation group on the $d$ circles by restricting the torus to be rectangular.
If three of the circles are much
larger than the others, the theory will behave like $3+1$ dimensional SYM at low energies and should 
have an approximate Montonen-Olive $SL(2,\Z)$ symmetry. In [\Alg], it was supposed that this should
extend to a symmetry for all values of the radii, at least at low energies, so that for any
3 of the $d$ circles, there is a Montonen-Olive transformation $S_{ijk}$. 
The transformation $S_{ijk}$ acts on the M-theory variables as [\Alg]
$$ 
R_i \to {l_p^3 \over R_jR_k },
\qquad
R_j \to {l_p^3 \over R_kR_i },
\qquad
R_k \to {l_p^3 \over R_kR_i },
\eqn\abc$$
with all other radii unchanged.
This is a T-duality combined with a permutation (from the Weyl group of $SL(d,\Z)$). The action of these transformations on the SYM variables is given in [\Alg].
The permutation group together with the Montonen-Olive transformations $S_{ijk}$ generate the Weyl group of 
$E_d(\Z)$ and these were used to build up U duality multiplets $E_d(\Z)$.
The momentum multiplet was generated by acting with U-duality on the Kaluza-Klein mode with momentum in the $i$'th direction
and energy
$$E={1\over s_i}
\eqn\abc$$ 
while the flux multiplet was obtained by acting with states carrying electric flux in the $i$'th direction and energy
$$E={g^2 s_i^2\over NV_s }
\eqn\abc$$ 
This gave rise to an infinite set of
  states, of which only a finite number are relevant for $d\le 8$.  
The first few states in the momentum and flux multiplets have
 the following masses and multiplicities
[\Alg]:
\vskip 1cm

\begintable
 Yang-Mills Energy | M-Theory Mass | Degeneracy \elt
 ${1\over s_i}$ | ${R_{11}R_i\over l_p^3}$ | $d$ \elt
 ${s_{i_1}\cdots s_{i_{d-4}}\over g^2}$ | ${R_{11}R_{j_1}\cdots R_{j_4}\over l_p^6}$ | 
$ {d! \over 4! (d-4)!}
$ \elt
 ${V_s\over g^4} {s_{i_1}\cdots s_{i_{d-6}}\over s_i}
$ | ${R_{11}R_i^2R_{j_1}\cdots R_{j_5}\over l_p^9}$ | 
$6{d! \over 6! (d-6)!}$ \elt
 ${V_s^2\over g^6}{s_{i_1}\cdots s_{i_{d-7}}\over s_is_js_k}$ | ${R_{11}R_i^2 R_j^2R_k^2 R_{j_1}
\cdots R_{j_4}\over l_p^{12}}$ | $35
{d! \over 7! (d-7)!}$ 
\endtable

\centerline{{\bf Table 1} Momentum multiplet states.}

\vskip .5cm

\begintable
 Yang-Mills Energy | M-Theory Mass | Degeneracy \elt
 ${g^2 s_i^2\over N V_s}$ | ${1\over R_i}$ | $d$ \elt
 ${V_s\over Ng^2s_i^2 s_j^2}$ | ${R_i R_j
\over l_p^3}$ | ${d! \over 2! (d-2)!}$ \elt
 ${V_s s_{i_1}^2\cdots s_{i_{d-5}}^2\over N g^6}
$ | ${R_{j_1}\cdots R_{j_5}\over l_p^6}$ | ${d! \over 5! (d-5)!}$ \elt
 ${V_s^3\over N g^{10}}{s_{i_1}^2\cdots s_{i_{d-7}}^2\over s_i^2}
$ | ${R_i^2R_{j_1}\cdots R_{j_6}\over l_p^9}$ | $7{d! \over 7! (d-7)!}$ \elt
 ${V_s^5\over N g^{14}}{s_{i_1}^2\cdots s_{i_{d-8}}^2\over s_i^2s_j^2s_k^2}
$ | ${R_i^2R_j^2R_k^2R_{j_1}\cdots R_{j_5}\over l_p^{12}}$ | $56{d! \over 8! (d-8)!}$ 
\endtable

\centerline{{\bf Table 2} Flux multiplet states.}

\vskip .5cm

The states in the second row of the flux multiplet table  carry magnetic flux
in the $i,j$ directions and arise in M-theory from wrapped membranes, while the states in
the second row of the momentum multiplet table are $d-4$ branes arising from SYM instantons [\Alg],
or from longitudinal M5-branes in the M-theory picture. The SYM interpretation of the other states
is not well understood.
The momentum and flux multiplets of states fit into   the following representations of $SL(d,\R)$ and
$E_d$ [\Alg]
\vskip .5cm

\begintable
 $d$  | $\hbox{Mom Multiplet:}\atop \hbox {$SL(d,\R)$ Reps}$ | $\hbox{Mom Multiplet:}\atop \hbox
{$E_d$ Reps}$
 | $\hbox{Flux Multiplet:}\atop \hbox {$SL(d,\R)$  Reps}$
 | $\hbox{Flux Multiplet:}\atop \hbox {$E_d$ Reps}$
 \elt
 3  | 3 | (3,1) | 3+3 | $(3',2)$ \elt
 4  | 4+1 | 5 | 4+6 | 10 \elt
 5  | 5+5 | 10 | 10+5+1 | 16 \elt
 6   | 6+15+6 | 27 | 6+15+6 | $27' $ \elt
 7  | 7+35+42+35+7 | $126\subset 133$ | 7+21+21+7 | 56 \elt
 8   | 8+70+168+... | $2160 \subset 3875$ | 8+28+56+56+56+28+8 | $240\subset 248$ 
\endtable

\centerline{{\bf Table 3}  $SL(d,\R)$ and $E_d$ Representations of Momentum and Flux mulitplets.}

\vskip .5cm

For $d=7,8$, the states obtained in this way do not appear to fit into complete representations of
$E_7,E_8$, suggesting that there should 
be extra states missing from this list if U-diality is to be a symmetry of the BPS spectrum. For
example, for the
$d=7$ momentum mulitiplet, the decomposition of the {\bf 133} of $E_7$ into $SL(7)$ representations
is $7+35+48+1+35+7 $ instead of the $7+35+42+35+7 $ in table 3. Thus the {\bf 42} should be
part of a {\bf 48+1} representation of $SL(7)$.
The {\bf 42}  corresponds to states with  mass 
$${V_R R_{11}
\over 
l_p^9}
{R_i \over  R_j}
\eqn\abc$$
for $i \ne j$. The extra 7 states must arise from states with the same mass formula, but with
$i =j$ for each of the seven values of $i=j$, as they must be related to those for
$i\ne j$ by $SL(7)$ transformations. This gives seven states, each with mass
 $${V_R R_{11}
\over 
l_p^9}
={R_{i_1}....R_{i_7} R_{11}
\over 
l_p^9}
\eqn\abc$$
Similarly, the $d=8$ fluxes multiplet has {\bf 56} states with masses given by
$${V_R  
\over 
l_p^9}
{R_i \over  R_j}
\eqn\abc$$
for $i \ne j$ which must in fact fit into a {\bf 63+1} of $SL(8)$. Including states of this form
with $i=j$ gives a further 8 states, each with mass
$${V_R  
\over 
l_p^9}
={R_{i_1}....R_{i_8} 
\over 
l_p^9}
\eqn\mate$$
Such \lq M8-brane' states 
were first suggested in [\Alg].  Similarly, the $d=8$ momentum multiplet is completed to
a {\bf 3875} of $E_8$.
That such extra states   complete the $E_7,E_8$ representations was also found by the
authors of [\Alg].\foot{This is contained in a Note Added to the version of [\Alg] to be published
in Nucl. Phys. B. I would like to thank David Kutasov for informing me of this.}

This will now be generalised to include the Nahm  transformations of [\Hv].
Again, if three of the circles are much
larger than the others, the theory will behave like $3+1$ dimensional SYM at low energies and should have in addition
the Nahm symmetry of [\Hv]. Assuming that for low energies this extends to a symmetry for all values
of the $s_i$, as was done above for
the Montonen-Olive   symmetry, we obtain  a further discrete transformation 
  for any three circles
that, when combined  with the transformations above in (the Weyl group of) $E_d$, generates (the Weyl group of) $E_{d+1}$.
Then repeating the analysis of [\Alg] but with the extra Nahm transformations,  we should find SYM
states fitting into  multiplets of $E_{d+1}$. While the Nahm transformations are complicated to
apply on the SYM variables and will not be given explicitly here, they act simply on the M-theory variables: they correspond to the
permutation swapping $R_{11}$ with $R_i$ for some $i$ [\Hv]. Thus they enlarge the  the Weyl group of
$SL(d)$ to that of $SL(d+1)$. Acting on the flux multiplet given in table 2 then gives extra states, which can be organised according to the power of $R_{11}$ in the M-theory mass.
There is one state with mass
$${1\over R_{11}}
\eqn\extras$$
which corresponds to the D0-brane  in the type IIA
string theory limit. The first few states with $M \propto R_{11}^2$  have   the following M-theory masses:
$$\eqalign{
& {R_{11}^2R_{j_1}\cdots R_{j_6}\over l_p^9},
\qquad
{R_{11}^2R_j^2R_k^2R_{j_1}\cdots R_{j_5}\over l_p^{12}},
\cr
&{R_{11}^2R_jR_k R_{j_1}^2\cdots R_{j_5}^2\over l_p^{15}},
\qquad {R_{11}^2R_i^3 R_{j_1}^2\cdots R_{j_6}^2\over l_p^{18}},...
\cr}
\eqn\extra$$
The   first of these gives the   D6-brane in the type IIA
string theory limit.
The first few states with $M \propto R_{11}^3$  have     masses:
$${R_{11}^3 R_{j_1}\cdots R_{j_8}\over l_p^{12}},
\qquad
{R_{11}^3 R_{k_1}\cdots R_{k_5}  R_{j_1}^2\cdots R_{j_3}^2\over l_p^{15}},\qquad
{R_{11}^3    R_{j_1}^2\cdots R_{j_7}^2\over l_p^{18}},....
\eqn\extrath$$
The first of these is the D8-brane.
There is  a state with mass linear in $R_{11}$, which corresponds to a longitudinal version of the 
8-brane 
\mate, with mass
$${R_{11} R_{j_1}\cdots R_{j_7}\over l_p^9}
\eqn\extrae$$
The flux multiplet is the set of all states with mass independent of $R_{11}$, generated by acting on a momentum state with mass $1/R_{i}$ with $E_d$ and the momentum
multiplet is the set of all states with mass linear in $R_{11}$, generated by acting on a longitudinal membrane state with mass $R_iR_{11}/l_p^3$ with $E_d$. Similarly, there is an $E_d$ multiplet generated by acting with $E_d$ on the D6-brane with mass
$${R_{11}^2R_{j_1}\cdots R_{j_6}\over l_p^9}\ek
The first few members of this multiplet are given in \extras-\extrae, and form a
singlet of $E_6$ for $d=6$, a {\bf 56} of $E_7$ for $d=7$ and a
{\bf 248} of $E_8$ for $d=8$. Similarly, there is a multiplet with mass
 proportional to $R_{11}^3$, the first few members of which are listed in \extrath, giving a singlet
in $d=7$ etc. There are further multiplets with mass proportional to $R_{11}^n$ for $n>3$ which
first contribute for $d=8$. The state \extras\ is an $E_d$ singlet which appears for all $d$.

These states can be combined with the  states from table 3  to have the right counting to fit into
the following representations of
$E_{d+1}$:

\vskip 1cm

\begintable
 d | $E_d$: Momentum | $E_d$: Flux | $E_d$: Extra | $E_{d+1}$: All \elt
 3 | (3,1) | $(3',2)$ | (1,1) | 10 \elt
 4 | 5 | 10 | 1 | 16 \elt
 5 | 10 | 16 | 1 | 27 \elt
 6 | 27 |$ 27' $| 1+1 | 56 \elt
 7 | 133 | 56 | 56+1+1+1 | 248 \elt
 8 | 3875 | 248 | 1+248+... |  
\endtable

\centerline{{\bf Table 4} Representations of $E_{d+1}$.}

\vskip .5cm

They must certainly fit into representations of the $SL(d+1)$ subgroup of $E_{d+1}$ which acts
naturally on $T^d\times S^1$ irrespective of whether the circle is null or space-like, and   they
also transform as representations of $E_d$.
For $d=3$, they fit into representations of $SL(5;\Z)$ (with the provisos given earlier). Then it
follows
   they must in fact transform as the 
representations of
$E_{d+1}$ in table 4.

Thus the BPS states of the $d+1$ dimensional SYM theory on $T^d\times \R$ fit into representations of
$E_{d+1}(\Z)$, and these should extrapolate to BPS states of the full matrix theory.
These correspond to BPS states of M-theory compactified on $T^{d+1}$ (with one null circle), which
includes some novel states for $d\ge 7$, some of which were discussed in [\Alg].
For M-theory compactified on $T^d$, the flux multiplet states of table 2
give M-theory states with masses
 $$ \eqalign{
&{1\over R_i},\qquad
{R_i R_j
\over l_p^3},\qquad
{R_{j_1}\cdots R_{j_5}\over l_p^6},\qquad
{R_i^2R_{j_1}\cdots R_{j_6}\over l_p^9}, 
\cr &
{ R_{j_1}\cdots R_{j_8}\over l_p^9},\qquad
{R_i^2R_j^2R_k^2R_{j_1}\cdots R_{j_5}\over l_p^{12}},\qquad
{R_i^3  R_{j_1}\cdots R_{j_8}\over l_p^{12}},
\cr &
{R_iR_j R_{j_1}^2\cdots R_{j_6}^2\over l_p^{15}},\qquad
{R_i^3 R_{k_1}\cdots R_{k_5}  R_{j_1}^2\cdots R_{j_3}^2\over l_p^{15}},\qquad
{R_i^3    R_{j_1}^2\cdots R_{j_7}^2\over l_p^{18}},....
\cr}
\eqn\dood$$
where $i=1,...,d$.
The list is infinite, but only a finite number of these states occur for $d\le 8$, and \dood\
contains all states occuring for $d\le 8$. We have seen that these flux multiplet states combine with
the momentum multiplet states and the extra states \extras-\extrae\ to give states fitting into
representations of
$E_{d+1}$ that are given by precisely the {\it same} list as in \dood, but with the indices running
over $d+1$ values
$i=1,...,d,11$ to include the longitudinal direction $R_{11}$. The  same list of masses   would be
obtained for the \lq flux multiplet' of states in the compactification of M-theory
on a {\it space-like } torus $T^{d+1}$. Indeed, the same list could have been generated by
acting with $E_{d+1}(\Z)$  on the states with $M=1/R_i$. Thus U-duality gives the complete multiplet
of M-theory states which includes the expected wrapped M-branes etc, together with some unexpected
new states for $d\ge 7$.

In the following, it will be 
useful to refer to a state with
mass dependence proportional to
$$ (R_{i_1}\cdots R_{i_p})^\aa (R_{j_1}\cdots R_{j_q})^\bb ...(R_{k_1}\cdots R_{k_p})^\gg
\eqn\mas$$
with $\aa<\bb <....<\gg$
as a $(p^\aa, q^\bb,....,r^\gg)$-brane  and, if $\aa=1$, to 
refer to a $(p^1, q^\bb,....,r^\gg)$-brane as a $(p, q^\bb,....,r^\gg)$-brane (i.e. to drop the
exponent $\aa=1$), while a $(p^1)$-brane with $p\ge 0$ will be referred to as a $p$-brane, in the
usual way. A state in M-theory  with mass dependence \mas\
will have a mass proportional to $l_p^{-a}$ where $a=\aa p
+\bb q +...+\gg r +1$. Such states will appear in a compactification on $T^d$ only for $d\ge
p+q+...+r$.
The states in the list \dood\ are then $(p^1, q^\bb,....,r^\gg)$-branes with
$(p^1, q^\bb,....,r^\gg)$ given respectively by
 $$ \eqalign{
&(-1),\qquad
2,\qquad
5,\qquad
(6,1^2), 
\cr &
8,\qquad
(5,3^2)
,\qquad
(8,1^3)
\cr &
(2,6^2)
,\qquad
(5,3^2,1^3)
,\qquad
(7^2,1^3),....
\cr}
\eqn\doodo$$

\chapter {The BPS Brane Spectrum of M-Theory and Type II String Theory}

We now turn to the interpretation of the BPS states \dood,\doodo\ in M-theory and in type II string
theory. We will start by considering M-theory compactified on a space-like torus $T^d$, 
which has 0-brane states given by \dood\ with $i=1,...,d$.
%and at the end consider the case of a null torus.
In the limit in which one of the radii becomes small, $R_s $ say, then a weakly 
coupled IIA string theory emerges with
string coupling $g_s$ given by
$$ g_s={R_s\over l_s}
\eqn\abc$$
where the string length is
$$l_s^2={l_p^3
\over R_s}
\eqn\string$$
The weakly coupled string limit is then obtained by taking $g_s \to 0$ while keeping $g_s$ fixed.
Different string theory limits emerge for different choices of $R_k$.

The states with $M=1/R_m$ are Kaluza-Klein modes (or pp-waves) carrying momentum in the $m$'th
direction. 
Such states with $m=1,...,d$ generated the flux multiplet in table 2, while the 
state with $M=1/R_{11}$ was the first  \lq extra' state added
in \extras\ to obtain $E_{d+1}$ representations. It is a state with 
longitudinal momentum, corresponding to a D0-brane of type IIA.
 2-brane states with $M=R_mR_n/l_p^3$ are membranes wrapped around the $m,n$ directions.
States with $M=R_{m_1}...R_{m_5}/l_p^6$ are wrapped 5-branes.
As is well known, in the type IIA limit, the 2-brane gives a string or D2-brane, the 5-brane gives a
5-brane or D4-brane and the pp-waves give pp-waves or D0-branes.

The $(6,1^2)$ states with
$$M={R_1\dots R_6 R_7^2 \over l_p^9}
\eqn\kma$$
were considered in [\Alg]; we give a related  interpretation here. These are Kaluza-Klein monopoles,
or G-branes of M-theory [\town,\GravDu]. A KK monopole or G6-brane in M-theory is the space-time
$\R^{6,1}\times TN$ where $TN$ denotes Taub-NUT space. This can be wrapped to give the solution
$\R \times T^6\times TN$ and depends on the 6 radii $R_1,\dots ,R_6$ of $T^6$ and the 
asymptotic radius $R_7$ of the 
$S^1$ fibre of $TN$, which is proportional to the NUT parameter $N$.
These states carry the charge discussed in [\GravDu] which occurs in the supersymmetry algebra, and
the mass is given by this charge. It is
  proportional to $NR_1\dots R_6 R_7$ [\GravDu], which gives \kma\ as $N \propto R_7$.

For the $(6,1^2)$ states with mass \kma, taking $R_7$ as the   radius $R_s$ in \string\ corresponding
to the string coupling
$g_s$ gives a state with mass
$$M={R_1\dots R_6   \over g_s l_s^7}
\eqn\abc$$
and is the D6-brane of the IIA theory [\Alg].
If $R_1$ corresponds to $g_s$, the state is a $(5,1^2)$-brane with
$$M={R_2,\dots R_6 R_7^2 \over g_s^2 l_s^8}
\eqn\abc$$
and is  the KK monopole or G5-brane of the IIA theory, related by T-duality to the NS 5-brane of the
IIB theory [\HT,\GravDu]. For $g_s$ corresponding to another direction, $R_8$ say, the mass is
$$M={R_1\dots R_6 R_7^2 \over g_s^3 l_s^9}
\eqn\thre$$
so that this is a $(6,1^2)$-brane of type IIA theory.
T-duality in the 
$R_8$ direction gives a decompactification 
of the dual coordinate, $R'_8 \to \infty$ and gives
a IIB-brane with
$$M={R_1\dots R_6 R_7  \over g_s^3 l_s^8}
\eqn\abc$$
This is the (1,0) 7-brane of the type IIB theory which is related by $SL(2,\Z)$ to the (0,1)
D7-brane.  The fact that this has $1/g^3
$ dependence was shown in [\StrMem], where
the $g$ dependence of branes in string theory was calculated. We now review   and 
elaborate on this.

In the IIB theory, consider a $p$-brane whose energy density is proportional to $g^{-a}$ at weak coupling;
for a D-brane with $p=1,3,5,7,9$, $a=1$, while for a NS 5-brane or G5-brane (KK monopole), $a=2$.
The IIB theory has 1,5,7 and 9 branes labelled by 2 charges $(p,q) $ which transform under
$SL(2,\Z)$ [\GravDu]. The (0,1) branes are D-branes, while the (1,0) 1-brane and 5-brane couple to the 
2-form in the Neveu-Schwarz sector. Then, under the $SL(2,\Z)$ duality that takes $g \to \ti g \equiv
1/g$, these transform to states with energy density  proportional to $\ti g^{-\ti a} $ where $\ti
a={(p+1)/2-a}$. For a D-brane, $\ti a={(p-1)/2}$ and the (0,1) D-brane is transformed into a (1,0)
brane. The D-string transforms into the fundamental string with $M\sim 1$, the 3-brane is invariant
and now has
$M\sim \ti g^{-1}$, the D5-brane transforms to the NS 5-brane with
$M\sim \ti g^{-2}$, the D7-brane transforms to the (1,0) 7-brane
with
$M\sim \ti g^{-3}$, and the D9-brane transforms to the (1,0) 9-brane proposed in [\GravDu]
with
$M\sim \ti g^{-4}$.
T-duality can then be used to relate these to branes of the IIA theory.

Consider next $(8,1^3)$-branes
 with
$$M={R_1\dots R_8 R_9^3\over l_p^{12}}
\eqn\abc$$
Choosing $R_9$ to be the string direction corresponding to $g_s$ gives a state with mass
$$M={R_1 \dots R_8  \over g_sl_s^{9}}
\eqn\abc$$
and is the D8-brane of type IIA.  
In [\GravDu], it was argued that the D8-brane must come from an M-theory brane which
 carries a 9-form charge occuring on the right-hand-side of
the 11-dimensional superalgebra, so that it is in some ways like a 9-brane.
We have thus learned that this M9-brane which gives
  the M-theory origin of the D8-brane is a $(8,1^3)$-brane of M-theory.
If instead $R_1$ is chosen as the string direction, one obtains a $(7,1^3)$-brane
with
$$M={ R_2\dots R_8 R_9^3\over  g_s^3 l_s^{11}}
\eqn\abc$$
while choosing the transverse string direction, $X^{10}$, gives a $(8,1^3)$-brane with
$$M={ R_1R_2\dots R_8 R_9^3\over  g_s^4 l_s^{12}}
\eqn\abc$$
The $(8,1^3)$-brane corresponds to the IIA  9-brane proposed in [\GravDu].

The $(5,3^2)$ states
 with
$$M={R_1\dots R_5 R_6^2 R_7^2 R_8^2\over l_p^{12}}
\eqn\abc$$
  first contribute for compactifications of M-theory on an 8-torus.
Formally, taking the string theory limit 
 corresponding to shrinking  
$R_1$
   gives a $(5,2^2)$-brane  with mass  proportional to $g^{-2}$,
 shrinking  
$R_6$
   gives a $(4,3^2)$-brane  with mass  proportional to
$g^{-3}$ and
 shrinking  
$R_9$
   gives a $(5,3^2)$-brane  with mass  proportional to
$g^{-4}$.

The 8-brane with mass
$$M={R_1\dots   R_8 \over l_p^{9}}
\eqn\abc$$
was proposed in [\Alg] and on reducing to string theory gives a 7-brane with tension proportional to
$g_s^{-2}l_s^8$ and an 8-brane with tension proportional to $g_s^{-3}l_s^9$ [\Alg].
The $(2,6^2)$-brane gives a $(2,6^2)$-brane, a $(1,6^2)$-brane and a $(2,5^2)$-brane of type IIA,
with masses proportional to $l_s^{-15} g_s^{-5}$, $l_s^{-14} g_s^{-4}$, and $l_s^{-13} g_s^{-3}$,
respectively. 

The reduction of other branes can be obtained similarly. 
The T-duality transformation
$$ R_i \to {l_s^2 \over R_i}, \qquad
g_s \to g_s {l_s  \over R_i}
\eqn\abc$$
can be used to relate the new type IIA states  above to states of the type IIB
theory. This doesn't change the dependence on $g_s$, but does change the dependence on the $R_i$
and the string length. 
It will be useful to augment the notation above  for string theory to include 
dependence on the string coupling, so that
a $(p^\aa, q^\bb,....,r^\gg;x)$-brane has mass proportional to
$$g_s^{-x} (R_{i_1}\cdots R_{i_p})^\aa (R_{j_1}\cdots R_{j_q})^\bb ...(R_{k_1}\cdots R_{k_p})^\gg
\eqn\abc$$
with $\aa<\bb <....<\gg$.

The $ (5,1^2;2)$-brane or IIA  KK monopole gives a (5;2)-brane which is the NS 5-brane, a
$(5,1^2;2)$-brane or type IIB KK monopole,  or a $(5,2^2;2)$-brane, depending on which dimension is
dualised.
The $(6,1^2;3)$-brane gives a $(7;3)$-brane, a $(5,2^2;3)$-brane and a $(6,1^2,1^3;3)$-brane,
depending on which dimension is dualised.
As mentioned above, the $(7;3)$-brane is related by S-duality to the $(7;1)$-brane or D7-brane, so
that these are   (1,0) and (0,1) 7-branes.
 The $(8,1^3;4)$-brane or IIA \lq NS 9-brane' gives a $(9;4)$-brane, a $(7,2^3;4)$-brane and a
$(8,1^3,1^4;4)$-brane, depending on which dimension is dualised.
The  $(9;4)$-brane is S-dual to a $(9;1)$-brane, which is the  NS 9-brane proposed in [\GravDu], and
these are the (1,0) and (0,1) 9-branes of [\GravDu].

For M-theory compactified to 3 or less dimensions, there are new states outside the usual
brane-scan, most of which do not yet have a string theory or supergravity interpretation.
M-theory compactified on $T^d$ has 0-branes in the expected representations of $E_d$ for $d\le 7$.
For $d<7$, the number of elementary 0-branes (i.e. the dimension of the charge lattice [\HT]) is
equal to the number of vector fields in the supergravity theory and all 0-branes arise from wrapped
M2 and M5 branes, together with pp-waves in the internal dimensions. For $d=7$, 
$D=4$ supergravity has 28 vector fields, but 0-branes can carry electric and/or magnetic charge
with respect to each, so the charge lattice is   56-dimensional.   49 elementary
0-branes arise from wrapped
M2 and M5 branes and pp-waves, while the remaining 7 arise from $(6,1^2)$-branes or KK monpoles on
$T^7$ (with seven different ways of choosing the Taub-NUT fibre among the 7 compact dimensions).
For M-theory on $T^8$, we have learned that the 0-branes should fit into the {\bf  248}
representation of $E_8$, and that there is a matrix model (or rather SYM)  construction of all these
states. Some of these arise from the expected pp-waves, and wrapped
M2,M5 and $(6,1^2)$-branes, but this is not enough to
complete the representation. In addition, one needs  8 wrapped 8-branes,
56 wrapped $(5,3^2)$-branes, 28 wrapped $(2,6^2)$-branes
and 8 wrapped $(7^2,1^3)$-branes.
The counting of 8-branes is mysterious, but was motivated earlier.
For $d=9$, the 0-brane spectrum is infinite, with BPS
states arising from the wrapped branes discussed above, plus an infinite set of generalisations.

It is interesting to ask whether these BPS states   arise as solutions of the supergravity theory.  
The web of duality relations linking these solutions gives a recipe for their construction, and in general these need not give asymptotically flat solutions (and so will not carry  charges occuring in the $D=11$ superalgebra).
For example, we have seen that the $(6,1^2;3)$-brane whose mass \thre\ has the unusual $g_s^{-3}$ dependence arises from the M-theory $(6,1^2)$-brane or KK monopole by compactifying  
one of the transverse directions and  identifying this with the string coupling constant.
Thus to construct the corresponding supergravity solution, one must take a periodic array of KK monopoles
and then identify one of the transverse directions so that the
solution is specified by a harmonic function on 
$\R^2\times S^1$ instead of the usual multi-Taub-NUT case which is specified by a harmonic function on 
$\R^3$.
Now T-dualising can give a $(7;3)$-brane, a $(5,2^2;3)$-brane or a $(6,1^2,1^3;3)$-brane of the type IIB theory, depending on which 
direction is dualised. The $(5,2^2;3)$-brane arises from T-dualising one of the directions inside the 6-dimensional world-volume, and the 
$(7;3)$-brane arises by restricting to solutions which are independent of a transverse direction, and compactifying and T-dualising this direction. 
The $(p,1^2)$-branes  correspond to KK monopole $p$-brane solutions with the $1^2$ corresponding to
the Taub-NUT
fibre direction. This suggests that $(p,q^2)$-branes could  correspond to spaces with a $T^q$
fibration.

A similar set of states arises for compactification on a $d$-torus which includes a null circle, and
hence for matrix theory, although the wrapped branes can have different properties depending on whether
the brane wraps the longitudinal direction or not, and 
in certain situations 
the SYM theory is modified to become SYM on a non-commutative torus [\doug].
For M-theory on $T^d\times S^1$, the 0-branes are arranged into various
 multiplets of $E_{d+1}$ with different $R_{11}$ dependence; these are the D0-brane \extras, the
flux multiplet in table 2, the momentum multiplet in table 1, the D6-brane multiplet in \extra, the
D8-brane multiplet in \extrath, plus others for $d\ge 8$. On taking the matrix theory large $R_{11}$
limit   [\Mat], all multiplets except the flux and 0-brane multiplet have masses that diverge with
$R_{11}$ and  are truncated out. The flux multiplet is in precisely the correct representation of
$E_{d}$ to give all the 0-branes of M-theory compactified on $T^d$. For example, the flux multiplet
for  M-theory on $T^7=T^6\times S^1$ (occuring in the $d=7$ row of table 4) is a {\bf 56} of $E_7$,
and in the large $R_{11}$ limit gives all 56 0-branes of SYM on $T^6\times S^1$, which split into a
flux multiplet, a momentum multiplet and a D0-brane singlet and D6-brane singlet (occuring in the
$d=6$ row of table 4). 

In [\humat], the question was raised as to whether there could be an extra hidden dimension in the matrix theories for $d=7,8$, in the same way as there is for $d=4$. In $d=4$, this comes about because there is a 
0-brane which combines with the 4 states with momentum in each of the four directions on $T^4$ to form a {\bf 5} of $SL(5)$.
On $T^d$, there is a $d$-momentum which combines with other 0-branes in the momentum multiplet to form a representation of $E_d$, and the question arises as to whether this includes a {\bf d+1} of
$SL(d+1)$ in the cases in which $SL(d+1)$ is a subgroup of $E_d$ (this is so for $d=4,7,8$); if so, this would suggest that the theory is naturally formulated on $T^{d+1}$ instead of $T^d$.
For $d=4$, as just stated, the momentum multiplet is a {\bf 5} of $E_4=SL(5)$ and the matrix theory is naturally formulated on $T^5\times 
\R$. For $d=5$, the momentum multiplet is a {\bf 10} of $E_5=SO(5,5)$, which splits into a ${\bf 5+5'}$ of $SL(5)$, corresponding to 5 momentum modes and 5 winding modes of a 6-dimensional string theory. For $d=6$, the 
 momentum multiplet is a {\bf 27} of $E_6$, which splits into a ${\bf 6+21}$ of $SL(6)$, corresponding to 5 momentum modes and 21 membrane wrapping modes on $T^6$.

For $d=7$, the 
 momentum multiplet is a {\bf 133} of $E_7$, which splits into a ${\bf 63 +35 +35'}$ of $SL(8)$, 
and hence to a ${\bf 48+1+7+7 +35 +35'}$ of $SL(7)$.
 The 7 momentum modes do not form part of a {\bf 8} of $SL(8)$, so
there does not seem to be a natural formulation on $T^8$.
The ${\bf  35 +35'}$ are naturally associated with 3-brane and 4-brane wrapping modes on $T^7$, 
the extra {\bf 7} could correspond to string or   6-brane wrapping modes, but the {\bf 48 } is rather mysterious, suggesting an object coupling to a mixed second-rank tensor $L^i{}_j$ on $T^7$, instead of the antisymmetric tensor to which a brane couples.
A wrapped $p$-brane with volume form $\ww _{i_1... i_p}$
has mass proportional to
$$\ww ^{i_1... i_p} R_{i_1}... R_{i_p}
\ek
and the number of distinct states is the number of components of the volume form,
while the  states that combine with the momentum here have mass  
$${V_R R_{11} \over l_p^9} L^i{}_j{R_i \over R_j}
\ek
and the number of such states is 
$48+1$, the number of components of the tensor $L^i{}_j$. These consist of  
$42$ wrapped KK monopoles plus $7$ 8-brane states.

For $d=8$, the 
 momentum multiplet is a {\bf 3875} of $E_8$, which splits into a ${\bf 80 +240+240'+1050+1050'+1215}$ of $SL(9)$.
The ${\bf 80 }$ splits into a ${\bf 63 +1+8+8}$ of $SL(8)$, containing the 8-momentum, which does not
combine with a 0-brane to give a 9-momentum, but again combines with   mysterious  tensor states. 
Progress in the cases $d\ge 7$ will clearly require a better   understanding of these 8-branes and other exotic branes.

\refout

\bye